\documentclass[prb,showpacs,twocolumn,ams,amsmath]{revtex4}
\usepackage{graphicx}
\usepackage{multirow}

\begin{document}
\title{{\it Ab initio} Investigation of Magnetic Transport Properties by Wannier Interpolation}
\author{Yi Liu, Hai-Jun Zhang, and Yugui Yao}\email{ygyao@aphy.iphy.ac.cn}
\affiliation{Institute of Physics and Beijing National Laboratory
for Condensed Matter Physics, Chinese Academy of Sciences, P. O. Box
603, Beijing 100190, China}

\begin{abstract}
We present an efficient {\it ab initio} approach for the study of
magnetic transport properties based on the Boltzmann equation with
the Wannier interpolation scheme. Within the relaxation time
approximation, band-resolved electric conductivity under a finite
magnetic field is obtained and the historical motion of the electron
wave packet in reciprocal space is determined. As a typical
application of this method, we have calculated the electric
conductivities of MgB$_2$ under finite magnetic fields. Multiband
characters for the individual bands are revealed, and the field
dependence of the conductivity tensor is studied systematically with
the field orientated parallel and normal to the $c$-axis,
respectively. The obtained historical motion is employed to simulate
directly the cyclotron motion in the extremal orbit and determine
the corresponding effective mass. Moreover, This approach is further
exploited to calculate the Hall coefficient in the low-field limit,
without the complicated computation for the second ${\mathbf k}$
derivative of the band.
\end{abstract}

\pacs{71.15.Dx, 71.18.+y, 75.47.-m}

\maketitle

\section{introduction}
Magnetic transport properties, such as magnetoresistance
(magnetoconductivity) and the Hall effect, which have been
investigated experimentally and theoretically for a long history,
are of great importance in the field of research for metals and
semiconductors. These properties are sensitively dependent on the
topology of the Fermi surface (FS). The corresponding semiclassical
interpretations were first given by Lifshitz {\it et
al.}\cite{Lifshitz} in the light of the Boltzmann equation with a
general collision operator. Subsequently they were further developed
for the studies of magnetoresistance\cite{Wilson,Ziman} and Hall
effect,\cite{Hurd} respectively.

The Boltzmann transport theory has functioned successfully in
previous studies for two-dimensional systems through analytical and
numerical solutions.\cite{Beenakker,PRB41Zhang,PRL76BPS,PRB57Menne}
Yet in real three-dimensional crystals, transport behaviors are
significantly determined by the complex FS, where models are not
much satisfactory and analytical solutions are formidable. {\it Ab
initio} study is then essential to provide the most comprehensive
and reliable investigation. To our knowledge of magnetoresistance,
so far works based on electronic structure calculation have been
rarely concerned with the numerical solution of the Boltzmann
equation for the band-resolved electric conductivity under the
finite magnetic field.\cite{HuanYang, epl181Monni} They were faced
with either heavy {\it ab initio} computation of properties on the
accurate FS, or unreliable results from the Shankland-Koelling-Wood
(SKW) interpolation scheme\cite{JCP67Koelling,SKWB38Pickett} in the
presence of band crossings close to the FS. On the other hand, in
the low-field limit the solution of the Boltzmann equation has been
extensively studied for the Hall effect through the power-series
expansion of the magnetic field.\cite{Jones} With the isotropic
relaxation time approximation, the Hall coefficient is explicitly
formulated for metals having cubic symmetry.\cite{Hurd} Following
this perturbation method, previous simulations were carried out
based on empirical tight-binding (TB) band structures,\cite{B45TB}
{\it ab initio} calculation combined with the SKW,\cite{B61SKW} and
the Wannier interpolation scheme.\cite{B75WFs} However, only cubic
metals were involved in these studies, and the second $\mathbf k$
derivative of the band was necessary.

In this paper, we present an {\it ab initio} approach combined with
the Wannier interpolation scheme for the solution of the Boltzmann
equation within the relaxation time approximation. Band structures
are obtained through the {\it ab initio} calculation, and Wannier
functions (WFs) are then constructed.\cite{Vanderbilt12} The
Runge-Kutta scheme is employed to determine the historical motion of
the electron wave packet, where the WFs are used to evaluate the
local velocities. The numerical integration with the tetrahedron
scheme\cite{Blochl} is adopted to deal with the sum over ${\mathbf
k}$-space. It is noticeable that the velocity evaluation with the
aid of the Wannier interpolation is more efficient (reliable) than
that using {\it ab initio} calculation ( the SKW
interpolation\cite{B75WFs}) at individual ${\mathbf k}$ points. We
have already applied this magnetoconductivity simulation to the
superconductor MgB$_2$\cite{Nature410} in our previous
Letter\cite{HuanYang} and the combination with the experimental
measurements played a key role in deriving electron-phonon
scattering times. In this work, we further study systematically the
band-resolved electric conductivity of MgB$_2$ based on this
approach, with the finite magnetic field orientated parallel and
normal to the $c$-axis, respectively. The historical tracks obtained
during the Runge-Kutta integration is also used to study directly
the cyclotron motion of the electron wave packet in the extremal
orbit. This provides a direct approach to determine the effective
mass of the corresponding extremal orbit. Moreover, we are able to
calculate the Hall coefficient in the low-field limit based on the
conductivity simulation, which is applicable for crystals of
arbitrary structures. Unlike the perturbation method, the
complicated computation of the second ${\mathbf k}$ derivative of
the band is avoided.

The paper is organized as follows. The simulation method is
presented in Sec. II. It is employed to study systematically the
magnetoconductivity of MgB$_2$ in the next section, where the
simulation of the cyclotron motion of the electron wave packet is
included. In Sec. IV, we calculate the low-field Hall coefficients
for several cubic metals and hcp Mg. Conclusions are drawn in
section V.

\section{simulation method}
With the powerful Wannier interpolation scheme,\cite{B74Wang,B75WFs}
we calculate the band-resolved electric conductivity under a finite magnetic
field through numerically solving the Boltzmann equation. In this
section, all the ingredients involved in our simulation are detailed
below.

\subsection{Band-resolved electric conductivity}
According to the Boltzmann transport theory within the relaxation
time approximation, the dc electric conductivity tensor of a metal
under a uniform magnetic field is written as\cite{SolidState}
\begin{eqnarray}
\mathbf\sigma^{(n)}=\frac{e^2}{4\pi^3}\;\int d\mathbf k\,
\tau_n(\varepsilon_n(\mathbf k))\mathbf v_n(\mathbf k)\bar{\mathbf v}_n(\mathbf k)
(-\frac{\partial f}{\partial\varepsilon})_{\varepsilon=\varepsilon_n(\mathbf k)}\, ,\nonumber\\
\end{eqnarray}
where $n$ is the band index, $\tau_n(\varepsilon_n(\mathbf k))$ is
the scattering time of the $n$-th band and assumed to be dependent
on the wave vector $\mathbf k$ only through the eigenvalue
$\varepsilon_n({\mathbf k})$. $f$ is the Fermi-Dirac distribution.
 $\mathbf v_n(\mathbf k)$ is the velocity given by the equation of motion
\begin{equation}
{\mathbf v}_n({\mathbf
k})=\frac{1}{\hbar}\frac{\partial\varepsilon_n({\mathbf
k})}{\partial{\mathbf k}}\,.
\end{equation}
$\bar{\mathbf v}_n(\mathbf k)$ is a weighted average of the velocity
over the past history of the electron passing through $\mathbf k$:
\begin{eqnarray}
\bar{\mathbf v}_n(\mathbf k)=\int^0_{-\infty}\frac{\mathrm d t}{\tau_n(\mathbf k)}e^{t/\tau_n(\mathbf k)}
\mathbf v_n(\mathbf k_n(t))\,.
\end{eqnarray}
The magnetic field is explicitly acting on the time evolution of
$\mathbf k_n(t)$, as given in the equation of motion:
\begin{equation}
\frac{\mathrm d\mathbf k_n(t)}{\mathrm dt}=-\frac{e}
{\hbar}\mathbf v_n(\mathbf k_n(t))\times\mathbf B\,,
\end{equation}
with the initial condition $\mathbf k_n(0)=\mathbf k$. We start with
this condition for one $\mathbf k$ point and simulate reversely the
time evolution given by Eq.~(4). It is physically reasonable and
mathematically maneuverable, and follows a typical Runge-Kutta
method to deal with the equations of motion. The velocity at an
arbitrary ${\mathbf k}$ points is evaluated analytically through the
Wannier interpolation,\cite{B75WFs} as detailed in the following
subsections. A serial of $\mathbf k(t)$ points is obtained as a
result of the Runge-Kutta integration for each initial $\mathbf k$
point. They sample the history track of the electron wave packet at
the corresponding time. Consequently, $\bar{\mathbf v}(\mathbf k)$,
according to Eq.~(3), can be obtained as the weighted average of the
velocities at these $\mathbf k(t)$ points along the track. Then the
integration in Eq.~(1) is performed by employing the modified
tetrahedron scheme.\cite{Blochl}

 As in most cases, the relaxation time for each band is not easily
 determined. Therefore, we incorporate the magnetic field $B$ and
 the relaxation time $\tau$ as one variable $B\tau$, which relates
 to an important dimensionless quantity $\omega\tau=\frac{e}{m^*}B\tau$,
 with $m^*$ the effective mass. Phenomenally, $\omega\tau$ represents
 the completed part in the cyclotron orbit before the carrier is scattered out.

An implicit rule holds in the semiclassical framework: the magnetic
field does not affect the kinetic energy of the carrier. Thus it
imposes a severe restriction on every point in the ${\mathbf k}(t)$
serials that they are all sampled from the same constant-energy
surface. It is a straightforward consequence from the theoretical
viewpoint, but not accomplished automatically in the numerical
simulation. In our code, we developed a modified self-adaptive
Runge-Kutta scheme, which takes much less time than the regular
Runge-Kutta methods to achieve the same accuracy. We did convincing
tests for several arbitrary initial $\mathbf k$ points and found
that the energies in their ${\mathbf k}(t)$ serials were converged
better than $10^{-7}$ Hartree.
\subsection{Wannier interpolation}
In our simulation, the velocity at an arbitrary $\mathbf k$ point is
evaluated through analytical calculation with the Wannier
interpolation scheme. WFs are constructed as a ``postprocessing''
operation based on Bloch eigenstates and eigenvalues obtained
through a standard {\it ab initio} electronic structure calculation
carried out on a uniform $k$-point grid.\cite{Vanderbilt12} Here we
employ the ultrasoft-pseudopotential\cite{ultrasoft} plane-wave
method with generalized-gradient approximation\cite{pbegga} for the
exchange and correlation potential. We give a brief review about the
Wannier interpolation as follows, while details can be found in
Ref.~\onlinecite{Vanderbilt12}. We use the symbol ${\mathbf q}$ to
denote the points on the {\it ab initio} mesh, and ${\mathbf k}$ for
arbitrary and interpolation-grid points.

The eigenstate $\vert\psi_{n{\mathbf q}}\rangle$ is obtained after
the electronic structure calculation. Its periodic part is defined
as $u_{n{\mathbf q}}({\mathbf r})=e^{-i{\mathbf q}\cdot{\mathbf
r}}\psi_{n{\mathbf q}}({\mathbf r})$. Denoting the Wannier function
in the cell ${\mathbf R}$ within the band $n$ as $\vert{\mathbf
R}n\rangle$, we have
\begin{subequations}\label{eq:WannBloch}
\begin{eqnarray}
&&\vert{\mathbf R}n\rangle=\frac{V}{(2\pi)^3} \int d{\mathbf q}
\,e^{-i{\mathbf q}\cdot{\mathbf R}}\vert\psi_{n{\mathbf q}}\rangle\,,\\
&&\vert\psi_{n{\mathbf q}}\rangle=\sum_{\mathbf R} e^{i{\mathbf q}\cdot{\mathbf R}}\vert{\mathbf R}n\rangle\,.
\end{eqnarray}
\end{subequations}
Here $V$ is the volumn of the real-space primitive cell. As shown in
Eq.~(5a), the freedom in the choice of the phases of the Bloch
states results in the arbitrariness of the Wannier functions. By
minimizing their total delocalization with respect to the phase
freedom, the unique set of maximally localized Wannier functions is
determined. The corresponding Bloch-like functions $\vert
u_{n{\mathbf q}}^{(W)}\rangle$ are then obtained according to
Eq.~(5b). The superscript $(W)$ or $(H)$ marks the quantities
belonging to the Wannier or Hamiltonian gauge,
respectively.\cite{B74Wang,B75WFs} Straightforwardly we obtain the
Hamiltonian by
\begin{equation}
H_{nm}^{(W)}({\mathbf q})=\langle u_{n{\mathbf
q}}^{(W)}\vert\hat{H}({\mathbf q})\vert u_{m{\mathbf
q}}^{(W)}\rangle\,.
\end{equation}
With a Fourier transform, we have
\begin{equation}
\langle {\mathbf 0}n\vert \hat{H}\vert {\mathbf
R}m\rangle=\frac{1}{N_q^3}\sum_{\mathbf q}e^{-i{\mathbf
q}\cdot{\mathbf R}}H_{nm}^{(W)}({\mathbf q})\,,
\end{equation}
where $\hat{H}$ is the effective one-particle Hamiltonian. The sum
runs over all the ${\mathbf q}$ points on the {\it ab initio} mesh.
Finally the interpolation is implemented by transforming back the
matrix $\langle {\mathbf 0}n\vert \hat{H}\vert {\mathbf R}m\rangle$
to an arbitrary ${\mathbf k}$ point,
\begin{equation}
H_{nm}^{(W)}({\mathbf k})=\sum_{\mathbf R}e^{i{\mathbf
k}\cdot{\mathbf R}}\langle{\mathbf 0}n\vert\hat{H}\vert{\mathbf
R}m\rangle\,.
\end{equation}
The Hamiltonian at any ${\mathbf k}$ point is then analytically determined.
\subsection{Evaluation of velocity}
To calculate the velocity, we diagonalize the above interpolated
Hamiltonian by finding a unitary rotation matrix $U({\mathbf k})$:
\begin{equation}
H^{(H)}({\mathbf k})=U^{\dagger}({\mathbf k})\,H^{(W)}({\mathbf
k})\,U({\mathbf k})\,,
\end{equation}
where $H^{(H)}_{nm}({\mathbf k})=\varepsilon_{n{\mathbf
k}}^{(H)}\delta_{nm}$, and $\varepsilon_{n{\mathbf k}}^{(H)}$ are
identical to the eigenvalues obtained from the {\it ab initio}
calculation at ${\mathbf k}={\mathbf q}$. We then evaluate the
velocities for an arbitrary ${\mathbf k}$ point through
\begin{eqnarray}
\mathbf v_n(\mathbf
k)=\frac{1}{\hbar}\frac{\partial\varepsilon^{(H)}_n(\mathbf
k)}{\partial{\mathbf
k}}=\frac{1}{\hbar}\frac{\partial}{\partial{\mathbf
k}}H^{(H)}_{nn}\,.
\end{eqnarray}
Note that the WFs are constructed via discrete Fourier transform
based on the {\it ab initio} band structure. The interpolation will
not be jeopardized by band crossings and avoided
crossings.\cite{B74Wang,B75WFs}

\section{magnetoconductivity under finite magnetic fields}
As the first remarkable application of our approach, we study the
electric conductivities of normal-state MgB$_2$ under finite
magnetic fields. This superconductor has been investigated
extensively in recent
years.\cite{PRL86Kortus,PRL87Liu,PRB66Eltsev,PRL96Li} In our
previous Letter\cite{HuanYang}, the simulation method presented in
Sec. II was employed and the combination with the experimental
measurements provided a unique way to obtain band-resolved
scattering rates in multiband systems. In this work, we further
study thoroughly the anisotropic magnetoconductivity of MgB$_2$ with
the magnetic field orientated parallel and normal to the $c$-axis,
respectively.

The FS of MgB$_2$ consists of four sheets. Two forming slightly
warped cylinders come from a couple of ``holelike''
quasi-two-dimensional bands (bonding $\sigma_1$ and antibonding
$\sigma_2$), and two tubular sheets form the three-dimensional $\pi$
bands (antibonding $\pi_1$ and bonding $\pi_2$), as shown in Fig.~1.
With the magnetic field along the $c$-axis, there are six extremal
orbits centered around $\Gamma$ and $A$, as labelled in Fig.~1 by
numbers 1-3 and 4-6, respectively. The coordinate system is set with
the $c$-axis along the $z$ direction throughout the conductivity
calculation.
\begin{figure}
\includegraphics[width=8cm]{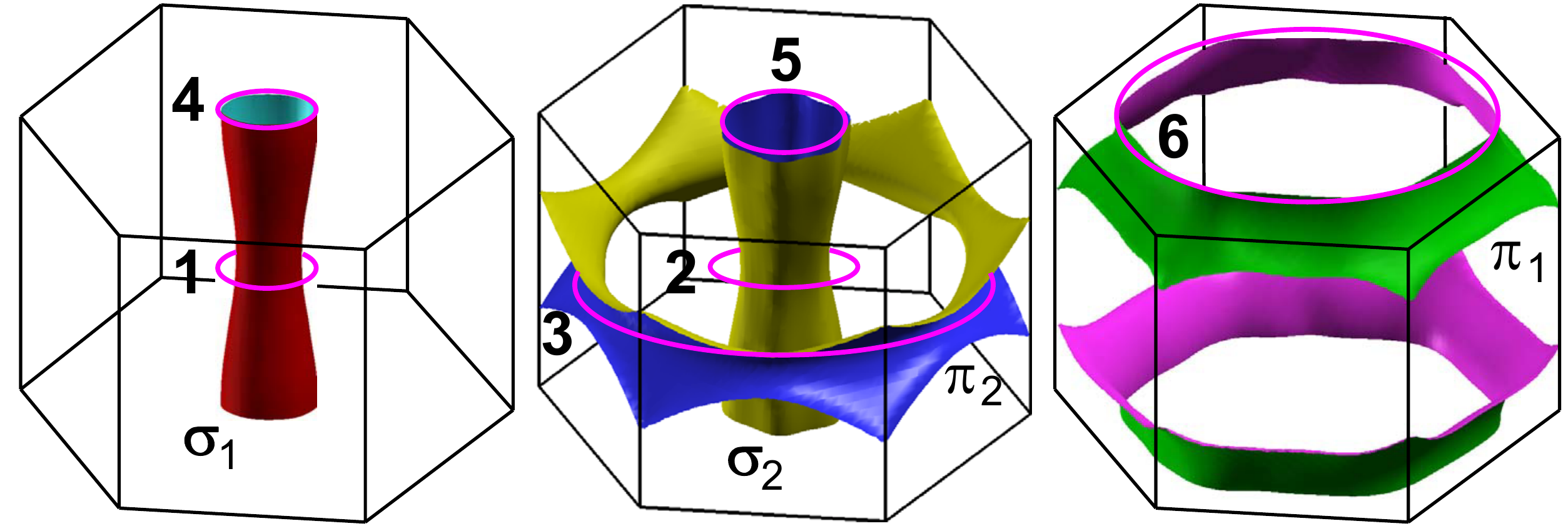}
\caption{(color online) The calculated Fermi surface of MgB$_2$ and
the extremal orbits labelled for the magnetic field along the
$c$-axis.}\label{fig1}
\end{figure}

Before calculating the magnetoconductivity of MgB$_2$, we first
analyze its asymptotic behavior for each band in the classical
high-field limit. According to the Boltzmann transport theory, the
change in electric conductivity under a finite magnetic field arises
from the cyclotron motion of the electron driven by the field. In
general, contributions from the cyclotron motions in a closed orbit
and an open orbit have different field dependences. Following the
analytical deduction by Lifshitz {\it et al.}\cite{Lifshitz}, and
the qualitative analysis by Pippard\cite{RepPippard} and
Fawcett\cite{AdvFawcett} based on the geometric features of the FS,
we can determine the asymptotic behavior of the diagonal element of
the electric conductivity tensor of MgB$_2$, as summarized in Table
1. It is helpful to check the validity of the following numerical
simulation under finite magnetic fields.

\begin{table}
\caption{\label{tab:table1} The asymptotic behaviors of the diagonal
elements of the electric conductivity tensor of MgB$_2$ in the
classical high-field limit. The coordinate system is defined with
the $c$-axis along the $z$ direction.}
\begin{ruledtabular}
\begin{tabular}{c c c c c c}
Field&Band&Orbit&$\sigma_{xx}$&$\sigma_{yy}$ &$\sigma_{zz}$ \\
\hline
\multirow{2}{*}{${\mathbf B}\parallel\hat{z}$} &$\sigma_1$, $\sigma_2$& Closed & $\sim B^{-2} $ &$\sim B^{-2} $ &$\sim B^{0}$\\
&$\pi_1$, $\pi_2$ & Closed &$\sim B^{-2} $ &$\sim B^{-2} $ &$\sim B^{0}$\\
\hline
\multirow{2}{*}{${\mathbf B}\parallel\hat{x}$} &$\sigma_1$, $\sigma_2$& Open &$\sim B^{0} $ &$\sim B^{0} $ &$\sim B^{-2}$\\
&$\pi_1$, $\pi_2$ & Closed &$\sim B^{0} $ &$\sim B^{-2} $ &$\sim
B^{-2}$
\end{tabular}
\end{ruledtabular}
\end{table}

\subsection{Magnetic field along the $c$-axis}

When the field is applied in the $c$ direction, all the four Fermi
sheets are composed of closed orbits. We use Eq.~(1) to calculate
the electric conductivity along the $x$-axis in the $ab$ plane.
$\sigma_{xx}/\tau$ of the four bands are shown in Fig.~2.
\begin{figure}
\includegraphics[width=8cm]{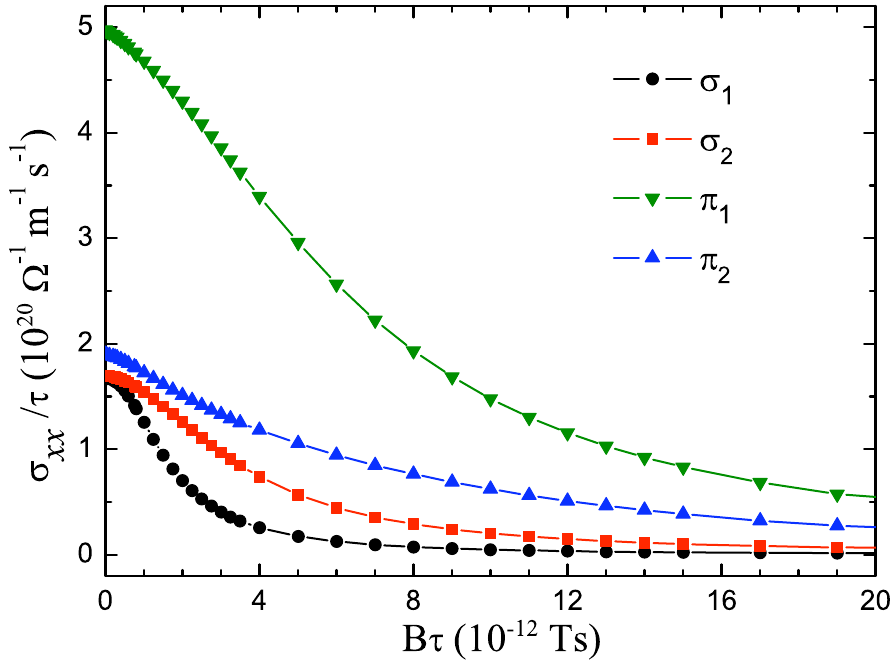}
\caption{(color online) $\sigma_{xx}/\tau$ as a function of $B\tau$
of the four bands of MgB$_2$, with the magnetic field along the
$c$-axis.} \label{fig2}
\end{figure}

The simulation is carried out with the FS refined after the
conventional electronic structure calculation.\cite{Blochl} With the
magnetic field along the 6-fold rotation axis, the symmetry of the
system are reduced by half, and $\sigma_{yy}$, which are not plotted
here, behave the same as $\sigma_{xx}$.\cite{Pippard} The field
dependence of the band-resolved $\sigma_{xx}$ can be described well
with a dominant term $B^{-2}$,\cite{HuanYang} in good agreement with
the asymptotic behavior in Table~1. Since the magnitudes of
$\sigma_{xx}/\tau$ of the two $\pi$ bands in Fig.~2 are noticeably
different, understanding of generalized $\sigma$ and $\pi$ bands
might be incomplete, and the individual bands deserve further
studies.

As stressed in Sec. II, the historical motion of the electron wave
packet in reciprocal space is obtained in the process of the
conductivity calculation. We take advantage of these cyclotron
motions in the extremal orbits of the four bands to determine the
corresponding effective masses. It provides a straightforward
approach to investigate the extremal orbit and effective mass, which
is usually a major concern of the study of the de Hass-van Alphen
(dHvA) effect.\cite{SolStaC121,PRL88Yelland,PRB65Mazin}

In one of these previous studies of the dHvA effect in MgB$_2$, the
calculation was carried out with
$m^*=\frac{\hbar}{2\pi}\oint\frac{d{\mathbf k}}{\vert{\mathbf
v}({\mathbf k})\vert}\,,$ and provided the absolute values of the
effective masses.\cite{SolStaC121} Later, Mazin {\it et
al.}\cite{PRB65Mazin} obtained the effective mass through the
standard formula: $m^*=\frac{\hbar^2}{2\pi}(\frac{\partial
\mathcal{A}}{\partial E})_{k_{z}}\,,$ with $\mathcal{A}$ the area of
the orbit, which was computed in small energy ranges around the
Fermi level.

In our case, the cyclotron motion of the electron wave packet at the
FS is tracked accurately with the serial of ${\mathbf k}(t)$ points
obtained through the self-adaptive Runge-Kutta integration for every
initial ${\mathbf k}$ point. Particularly, in the extremal orbits,
as labelled in Fig.~1, we choose one initial ${\mathbf k}$ point
each to study the historical motion. As an example, we plot in
Fig.~3 two of these historical tracks in Orbit 1 and Orbit 3,
respectively.

\begin{figure}
\includegraphics[width=8cm]{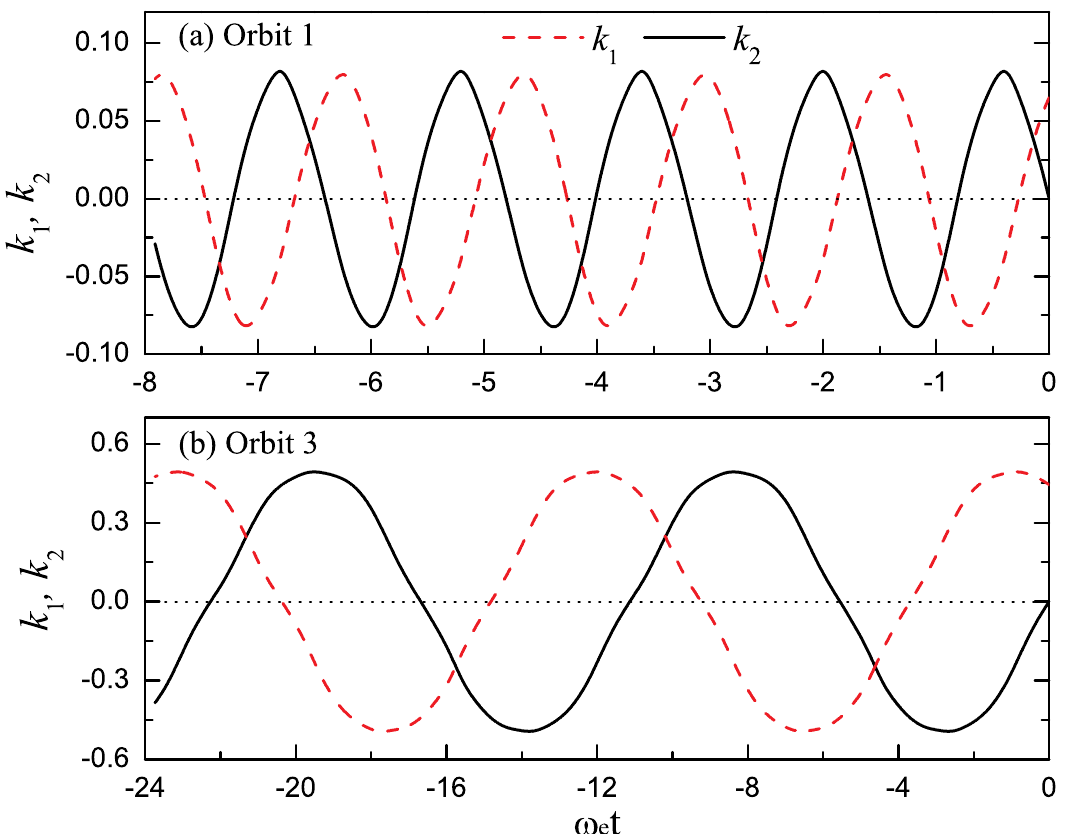}
\caption{(color online) Historical track of the electron wave packet
in reciprocal space. The initial positions are ${\mathbf k}(0.065,
0.0, 0.0)$ in Orbit 1 (a), and ${\mathbf k}(0.444, 0.0, 0.0)$ in
Orbit 3 (b), respectively. ${\mathbf k}$ points are expressed as
${\mathbf k}(k_1, k_2, k_3)$, in the unit of reciprocal lattice
vectors. There are $4000$ points on each curve. $k_3$ is fixed in
the $\Gamma$ plane during the simulation. The abscissa $\omega_e t$
is related to $Bt$ as $\omega_e t=\frac{e}{m_e}Bt$, where $m_e$ is
the mass of a free electron. }\label{fig3}
\end{figure}

These sinusoidal-like curves shown in Fig.~3 illustrate vividly the
cyclotron motion in reciprocal space driven by the magnetic field.
The period of the motion, $\mathcal{T}$, can be read directly from
the figure. The effective mass of the corresponding extremal orbit
is then straightforwardly obtained in the absolute value as $\vert
m^*\vert/m_e=\mathcal{T}/2\pi$, where $m_e$ is the free-electron
mass.
\begin{table}
\caption{\label{tab:table2} Calculated effective masses in the
extremal orbits of MgB$_2$, given in the unit of the free-electron
mass.}
\begin{ruledtabular}
\begin{tabular}{ccccccc}
Orbit: & 1  &2 & 3 & 4 & 5 & 6\\
\hline
Ref. \onlinecite{SolStaC121} & 0.254  & 0.313  & 1.699 & 0.550  & 0.612  & 0.924 \\
Ref. \onlinecite{PRB65Mazin} & -0.251 & -0.543 & 1.96  & -0.312 & -0.618 & -1.00 \\
This work  & -0.255 & -0.541 & 1.816 & -0.309 & -0.656 & -0.988\\
\end{tabular}
\end{ruledtabular}
\end{table}
Then the sign of the effective mass is determined through analyzing
the direction of the cyclotron motion. As shown in Fig.~3, the
cyclotron motion in Orbit 3 is anticlockwise, while it is clockwise
in Orbit 1, as well as those in the rest four extremal orbits. It is
then concluded that the third orbit is ``electronlike'', while the
rest of all are ``holelike''. Our results, listed in Table~2, show
good agreement with the previous studies, which confirms the
validity of the approach.

In the rest of this subsection, we return to the investigation of
band-resolved electric conductivity. In the current geometry, the
Hall conductivity, $\sigma_{xy}$, is another non-trivial element
besides $\sigma_{xx}$. In our previous Letter,\cite{HuanYang} the
Hall conductivity was calculated with Eq.~(1) for all the four
bands. A sign change from positive to negative has been found in the
$\pi_2$ band with increasing field, while the $\pi_1$ and the two
$\sigma$ bands behave simply as holes. For a thorough understanding,
we further perform the partial integration for the Hall
conductivity, $\sigma_{xy}^{(n)}(k_z)/\tau^{(n)}$, by
\begin{eqnarray}
\sigma_{xy}^{(n)}(k_z)/\tau^{(n)}&=&\frac{e^2}{4\pi^3}\int
\mathrm{d}k_x\mathrm{d}k_y\int_{-k_z}^{k_z}\mathrm{d}k_zv^{(n)}_{x}({\bf
 k})\bar{v}_y^{(n)}(\mathbf{k})\nonumber\\
&&\times(-\frac{\partial
  f}{\partial\varepsilon})_{\varepsilon=\varepsilon_n(\mathbf{k})}\,.
\end{eqnarray}
We employ a $30\times30\times40$ mesh, which gives $3192$ $ {\mathbf
k}$ points in the irreducible Brillouin zone. The extra finer grid
along the $z$-axis is adopted for the integration from $-k_z$ to
$k_z$. $B\tau$ is fixed at $1.0\times10^{-12}\;$Ts.

\begin{figure}
\includegraphics[width=8cm]{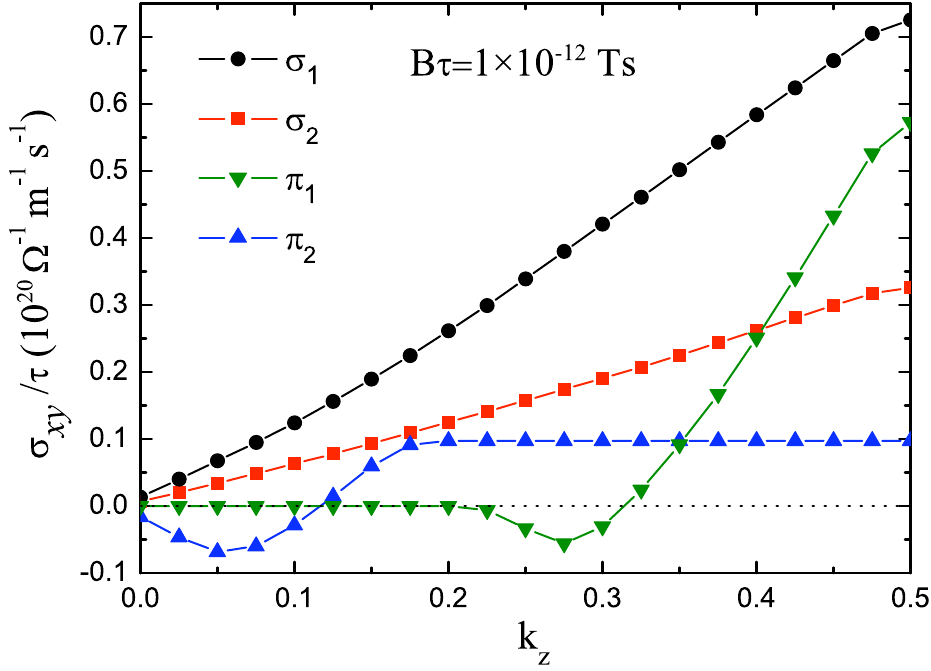}
\caption{(color online) The partial Hall conductivity
$\sigma_{xy}^{(n)}(k_z)/\tau^{(n)}$ contributed by orbits between
$-k_z$ to $k_z$ of the individual bands as a function of
$k_z$.}\label{fig4}
\end{figure}
The calculated partial Hall conductivity of the four bands is shown
in Fig.~4. For the two $\sigma$ bands, the linear growth is
understood with the cylindrical shape of the Fermi sheets. For the
two $\pi$ bands, there are sign changes along the $k_z$ axis, which
confirm that multiband character exists in the individual $\pi$
bands. Although the $\pi_1$ band was considered ``electronlike''
according to the band structure, ``holelike'' part occupies most of
the Fermi sheet except for the small peaks close to the $\Gamma$
plane. Thus it behaves as a hole in the full Hall conductivity with
the increasing magnetic field.\cite{HuanYang} Yet while the
situation is opposite for the $\pi_2$ band. The small hills are
``holelike'', and the big belly is ``electronlike''. Thanks to the
{\it ab initio} simulation of band-resolved conductivity, we have
obtained such detailed knowledge of the multiband character for the
individual bands in MgB$_2$.

\subsection{Magnetic field normal to $c$-axis}
When the magnetic field is normal to $c$-axis, different transport
behaviors are expected due to the hexagonal structure of MgB$_2$. We
further study this geometry by applying the field along the $x$-axis
in the $ab$ plane. The one-dimensional infinity of the two $\sigma$
bands is then exposed, and the tubular $\pi$ bands consist of closed
orbits in the planes normal to the field, including extended closed
orbits.

We calculate the electric conductivity tensor using Eq.~(1).
Symmetry is further reduced because the 6-fold rotation axis has
vanished in the current geometry. The three diagonal elements of the
conductivity tensor are shown in Fig.~5.
\begin{figure}
\includegraphics[width=8cm]{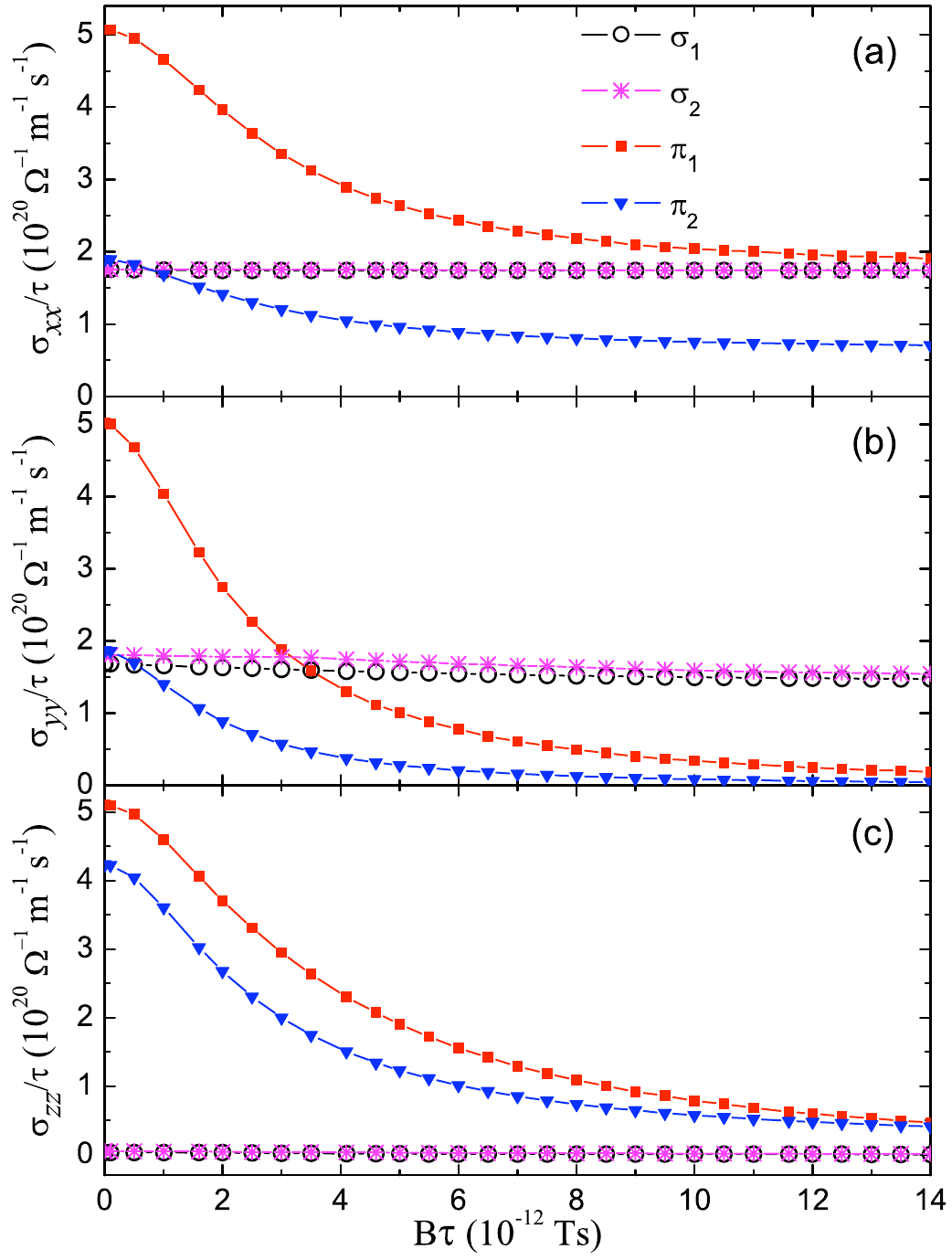}
\caption{(color online) $B\tau$ dependences of $\sigma_{xx}/\tau$
(a), $\sigma_{yy}/\tau$ (b), and $\sigma_{zz}/\tau$ (c) of the four
bands of MgB$_2$. The magnetic field is applied along the $x$-axis
in the $ab$ plane.}\label{fig5}
\end{figure}

There are two categories of field-dependent behaviors of the
diagonal elements. For the two $\sigma$ bands, conductivity elements
show little dependence on the magnetic field. In the $x$ direction,
the motion of the electron wave packet is not affected by the field,
rendering $\sigma_{xx}$ field-independent. Similar behavior in the
$y$ direction is accounted for by open orbits in the two (nearly
standard) cylinders.\cite{Lifshitz,RepPippard} The $z$-components of
Fermi velocities are negligible compared to those in the other two
directions. This provides an elegant interpretation for the almost
vanished elements $\sigma_{zz}$. The two $\pi$ bands are composed of
closed orbits with the magnetic field along the $x$-axis. As the
field increases, $\sigma_{xx}$ of the two $\pi$ bands decrease and
tend to saturation in the high-filed limit, while $\sigma_{yy}$, as
well as $\sigma_{zz}$, shows a decay which can be described well by
a dominant term $B^{-2}$. All the results agree with the high-field
asymptotic behaviors listed Table~1. In addition, these conductivity
elements under finite fields can be fitted to obtain global field
dependences, as carried out in Ref. \onlinecite{HuanYang}.

In principle, there are two applications of the calculated
band-resolved electric conductivity under the finite magnetic
fields. With the determined relaxation time in each band,
magnetoresistance study can be performed.\cite{epl181Monni}
Alternatively, combining the calculation results with the
experimental measurements of conductance or resistance in different
temperatures can lead to quantitative estimation about these
scattering rates, including the impurity scattering and the
electron-phonon coupling.\cite{HuanYang}
It is worth pointing out that  the efficient calculation provides
reliable results and the simulation is performed at the accurate
Fermi level. We suggest that this {\it ab initio} approach is
helpful in the study of finite-field magnetic transport properties.

\section{Low-Field Hall Coefficient calculation}
Beyond the electric conductivity simulation under a finite magnetic
field, we further applied the method in the low-field limit, where
Hall effect is the major interest of the magnetic transport
research. Previous studies\cite{B45TB,B61SKW,B75WFs} followed the
perturbation method, which adopts the power-series expansion of the
magnetic field and gives the Hall coefficient for cubic crystals as
\begin{equation}
R_H=\sigma_H/\sigma_0^2\,.
\end{equation}
Here $\sigma_0$ is the conductivity:
\begin{eqnarray}
\sigma_0=\frac{e^2}{3\hbar^2}\sum_{\mathbf k}\,\tau({\mathbf k})\,\left[\,\mathbf{\nabla}_{\mathbf k}\varepsilon({\mathbf k})\,\right]^2\,\left[\,-\frac{\partial f(\varepsilon)}{\partial\varepsilon}\,\right]\,,
\end{eqnarray}
and $\sigma_H$ is the Hall conductivity:
\begin{eqnarray}
\sigma_H&=&\frac{e^3}{12}\sum_{\mathbf k}\tau^2({\mathbf k})\,{\mathbf v}({\mathbf k})\,\left[\,{\rm Tr}({\mathbf M}^{-1})-{\mathbf M}^{-1}\,\right]\,{\mathbf v}({\mathbf k})\nonumber\\
&&\times\left[\,-\frac{\partial f(\varepsilon)}{\partial\varepsilon}\,\right]\,.
\end{eqnarray}
${\mathbf M}^{-1}$ is the inverse mass tensor, and its elements are defined by
\begin{equation}
{\mathbf M}^{-1}_{\alpha\beta}=\frac{1}{\hbar^2}\frac{\partial^2\varepsilon}{\partial k_\alpha \partial k_\beta}\,.
\end{equation}
In this perturbation method, the evaluation of the second ${\mathbf
k}$ derivative of the band is a difficult task. It either gives
unreliable results with the SKW interpolation in the presence of
band crossings and/or near degeneracies,\cite{B61SKW} or follows a
very complicated formula with the Wannier
interpolation.\cite{B75WFs} For instance, Fig.~6 shows the complex
band structure of fcc Pd. It has three energy bands across the Fermi
energy, and multiple band crossings in the proximity of the Fermi
level. Expensive computation is inevitable during the SKW
interpolation due to the complexity of the band structure, and the
accuracy of the result will be threatened because of the band
crossings. Even with the Wannier interpolation, the heavy
computation of the inverse mass tensor still can not be avoided.
\begin{figure}
\includegraphics[width=8cm]{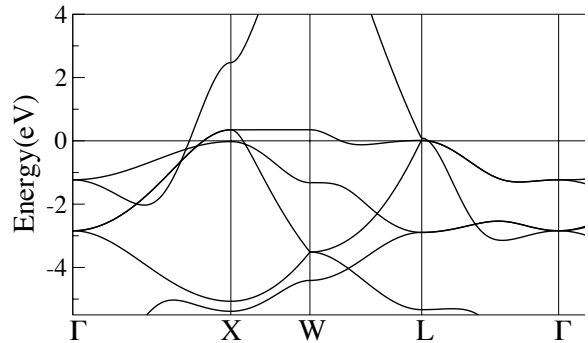}
\caption{The band structure of fcc Pd. The Fermi level is fixed at 0 eV.}\label{fig6}
\end{figure}

In our case, each element of the electric conductivity tensor can be
obtained through Eq.~(1). We can further evaluate the Hall
resistivity according to the definition directly:\cite{Pippard}
\begin{eqnarray}\label{eq:rho1}
 \rho_{yx}=\frac{-\sigma_{yx}}{\sigma_{xx}\sigma_{yy}-\sigma_{xy}\sigma_{yx}}\,.
\end{eqnarray}
Here we assume that the magnetic field is applied along the
$z$-axis. The low-field limit is achieved with $\rho_{yx}$ growing
linearly
 with the field. The Hall coefficient is then given by the definition, as well:\cite{Hurd}
\begin{equation}\label{eq:RH}
R_H=\frac{\rho_{yx}(B)-\rho_{xy}(-B)}{2B}\,.
\end{equation}
This scheme is valid for metals with arbitrary structures.
Particularly, in a crystal with threefold or fourfold symmetry about
the magnetic field direction, rotation transformations require
$\sigma_{xx}=\sigma_{yy}$ and $\sigma_{xy}=-\sigma_{yx}$. Thus we
have
\begin{equation}\label{eq:rho2}
\rho_{yx}
=\frac{\sigma_{xy}}{\sigma_{xx}^2+\sigma_{xy}^2}\,
\end{equation}
to simplify the calculation in the application to all cubic and
hexagonal systems. Throughout the calculation for the low-field Hall
coefficient, we do not need to deal with the inverse mass tensor.
The complexity of the band structure will not result in expensive
computation.

We performed the calculation of the low-field Hall coefficient with
Eqns.~(\ref{eq:rho1}), (\ref{eq:RH}), and (\ref{eq:rho2}) for typical cubic metals Li, Al, Cu, and Pd, as well as hcp Mg, where the magnetic field is along the hexagonal axis.
\begin{table}
\caption{\label{tab:table3}The calculated and experimentally
observed Hall coefficient $R_H$, in the unit of $10^{-11} {\rm
m}^3/{\rm C}$. Experimental data are all from Ref.
\onlinecite{Hurd}.}
\begin{ruledtabular}
\begin{tabular}{cccccc}
 & Li & Al & Cu & Pd & Mg\\
\hline
Ref.~\onlinecite{B45TB} & -12.8 & -1.7 & -5.2 & -6.0 & - \\
Ref.~\onlinecite{B61SKW} & -12.4 & -3.4 & -4.9 & -17  & - \\
Ref.~\onlinecite{B75WFs} & -12.7 & -2.5 & -4.9 & -11.9& -  \\
This work & -13.0 & -2.2 & -5.2 & -11.2 & -10.4 \\
Experiment& -15.5 & -3.43 & -5.17 & -7.6 & -8.3\\
\end{tabular}
\end{ruledtabular}
\end{table}
The results are listed in Table~3. For Li, Al, and Cu, we find great
agreement with previous simulations based on band structures fitted
through TB scheme,\cite{B45TB} {\it ab initio} calculation combined
with SKW interpolation\cite{B61SKW} and Wannier
interpolation.\cite{B75WFs} With a certain ratio between the
relaxation times of the two bands of Al\cite{B45TB}, we can
reproduce the experimental value in our simulation. In the case of
Pd, the result with the SKW interpolation is not as satisfactory as
those through TB and with the Wannier interpolation. It results from
the complex band structure as analyzed above. Our result agrees
excellently with that from Ref.~\onlinecite{B75WFs}, yet without the
complicated evaluation of the inverse mass tensor. The discrepancy
between calculated and experimental values was accounted for by the
constant relaxation time approximation.\cite{B45TB,JPF13Beaulac} For
hcp Mg, our result is comparable to the experimental data with the
magnetic field parallel to the hexagonal axis.
\section{conclusions}
We have developed an efficient numerical scheme for the solution of
the Boltzmann equation within the relaxation time approximation to
obtain band-resolved electric conductivity under a finite magnetic
field. It is based on the {\it ab initio} electronic structure
calculation, and the Wannier interpolation scheme is employed to
analytically evaluate the velocities along the historical track of
the electron wave packet. The simulation is guaranteed to be
performed at the accurate FS by the modified self-adaptive
Runge-Kutta method. As the first application, we have studied
systematically the electric conductivity of normal-state MgB$_2$
with the magnetic field orientated parallel and normal to the
$c$-axis, respectively. The multiband characters within the
individual bands are then revealed in detail.

During the conductivity calculation, the cyclotron motion of the
wave packet driven by the magnetic field is simulated. We have taken
advantage of these motions in the extremal orbits of MgB$_2$ to
calculate the corresponding effective masses, with the results in
good agreement with previous studies. It provides a direct
simulation of the cyclotron motion and a reliable evaluation of the
effective mass of the extremal orbit.

Another application of this method is the calculation of the
low-field Hall coefficient as a universal approach for metals and
semiconductors of arbitrary structures. We have reproduced the Hall
coefficients for several typical cubic metals, and for hcp Mg as an
example of non-cubic systems.

A remarkable appeal of this approach is that it provides an {\it ab
initio} method to simulate the electric conductivity under a finite
magnetic field. It is efficient and reliable owning to the Wannier
interpolation scheme. Moreover, the direct simulation of the
cyclotron motion is an elegant and useful application. The
calculation of the low-field Hall coefficient is universal for
crystals of arbitrary structures, and avoids all complicated
computation of the second derivatives of the band. Therefore, this
approach can be generally applied to a wide variety of systems for
magnetic transport study.

\section*{ACKNOWLEDGEMENTS}
We are grateful to Professor Junren Shi, Professor Haihu Wen and Dr.
Huan Yang for helpful discussions of transport properties of
MgB$_2$. This work is supported by the Natural Science Foundation of
China~(Nos.10674163, 10534030), the MOST Project~(Nos. 2006CB921300,
2007CB925000), the Knowledge Innovation Project of Chinese Academy
of Sciences.

\end{document}